\documentclass[conference]{IEEEtran}
\IEEEoverridecommandlockouts
\usepackage{cite}
\usepackage{amsmath,amssymb,amsfonts}
\usepackage{graphicx}
\usepackage{caption}
\usepackage{subcaption}
\usepackage{textcomp}
\usepackage{xcolor}
\usepackage{makecell}

\usepackage{algorithm}
\usepackage{algorithmicx, algpseudocode}
\usepackage{algpseudocode}

\usepackage[capitalize]{cleveref}

\newcommand{\E}{\mathbb{E}}
\newcommand{\C}{\mathbb{C}}

\newcommand{\cCN}{\mathcal{CN}}
\newcommand{\cN}{\mathcal{N}}
\newcommand{\bI}{\mathbf{I}}
\newcommand{\bH}{\mathbf{H}}
\newcommand{\bN}{\mathbf{N}}

\newcommand{\bU}{\mathbf{U}}
\newcommand{\bV}{\mathbf{V}}
\newcommand{\bW}{\mathbf{W}}
\newcommand{\R}{\mathbb{R}}
\newcommand{\bS}{\mathbf{S}}
\newcommand{\bSh}{\hat{\mathbf{S}}}
\newcommand{\bX}{\mathbf{X}}
\newcommand{\bY}{\mathbf{Y}}
\newcommand{\bZ}{\mathbf{Z}}
\newcommand{\bZh}{\hat{\mathbf{Z}}}
\newcommand{\bn}{\mathbf{n}}
\newcommand{\bu}{\mathbf{u}}

\newcommand{\bx}{\mathbf{x}}

\newcommand{\by}{\mathbf{y}}
\newcommand{\bz}{\mathbf{z}}

\newcommand{\bSigma}{\boldsymbol{\Sigma}}

\newcommand{\btheta}{{\boldsymbol{\theta}}}
\newcommand{\bepsilon}{{\boldsymbol{\epsilon}}}
\newcommand{\bphi}{{\boldsymbol{\phi}}}
\newcommand{\bvarphi}{{\boldsymbol{\varphi}}}

\def\BibTeX{{\rm B\kern-.05em{\sc i\kern-.025em b}\kern-.08em
    T\kern-.1667em\lower.7ex\hbox{E}\kern-.125emX}}
\begin{document}

\title{DM-MIMO: Diffusion Models for Robust Semantic Communications over MIMO Channels\\
}



\author{Yiheng Duan, Tong Wu, Zhiyong Chen and Meixia Tao \\
Cooperative Medianet Innovation Center, Shanghai Jiao Tong University, Shanghai, China\\
Emails: \{duanyiheng, wu\_tong, zhiyongchen, mxtao\}@sjtu.edu.cn
\thanks{This work is supported by the NSF of China under grant 62125108 and 62222111.}
}

\maketitle

\begin{abstract}
This paper investigates robust semantic communications over multiple-input multiple-output (MIMO) fading channels. 
Current semantic communications over MIMO channels mainly focus on channel adaptive encoding and decoding, which lacks exploration of signal distribution. To leverage the potential of signal distribution in signal space denoising, we develop a diffusion model over MIMO channels (DM-MIMO), a plug-in module at the receiver side in conjunction with singular value decomposition (SVD) based precoding and equalization. Specifically, due to the significant variations in effective noise power over distinct sub-channels, we determine the effective sampling steps accordingly and devise a joint sampling algorithm.
Utilizing a three-stage training algorithm, DM-MIMO learns the distribution of the encoded signal, which enables noise elimination over all sub-channels. 
Experimental results demonstrate that the DM-MIMO effectively reduces the mean square errors (MSE) of the equalized signal and the DM-MIMO semantic communication system (DM-MIMO-JSCC) outperforms the JSCC-based semantic communication system in image reconstruction.

\end{abstract}


\section{Introduction}
Recently, semantic communications have attracted extensive attention thanks to their great potential in improving transmission efficiency. By leveraging the rapid advancements in deep learning, semantic communications can adeptly extract and transmit meaningful semantic information through neural network (NN) based joint source-channel coding (JSCC), and have demonstrated superiority over traditional bit communications in various types of source transmissions \cite{10328187, 9398576, 9953110}. Thus far, semantic communications are regarded as a highly promising technique for 6G wireless communication networks and beyond \cite{ZHANG202260}.

Despite the great potential of semantic communications, most existing works primarily focus on single-input single-output (SISO) channels. It is thus of great importance and need to investigate semantic communications over multiple-input multiple output (MIMO) channels, given that MIMO has played a leading role in boosting channel capacity and transmission reliability since 3G wireless communications. The main distinction between SISO and MIMO channels in semantic communications lies in how to allocate semantic information over sub-channels with the aid of channel state information (CSI).
To this end, in \cite{wu2023deep}, the proposed DeepJSCC-MIMO adopts SVD-based precoding and equalization, constructing a channel-condition-based heatmap as an additional input for encoding and decoding. In \cite{zhang2023scan}, in addition to SVD-based precoding and equalization, channel and feature attention (CFA) modules are embedded in the JSCC encoder and decoder to adapt to MIMO channel conditions. 
However, despite the above channel adaptive encoding and decoding, performing signal denoising also holds the potential to further enhance the performance of semantic communications under MIMO channels, which requires further investigation.
 
As an advanced type of generative models, diffusion models (DMs) not only achieve great success in image generation \cite{song2022denoising} but also show advancements in image \cite{yilmaz2023high} and audio \cite{grassucci2023diffusion} restoration. 
Introducing information decay to source data in the forward diffusion process  by adding noise, DMs are trained to learn the decay with NNs.
With noise of different power captured in different steps, DMs are not only capable of sample generation but also available for signal denoising.
For example, recently, channel denoising diffusion models (CDDM) are proposed to mitigate the impact of channel noise in SISO channels with the adaptive forward diffusion and the corresponding reverse sampling process \cite{wu2023cddm}.

\begin{figure*}[t]
    \vspace{-1em}
    \centering
    \includegraphics[width=0.8\textwidth]{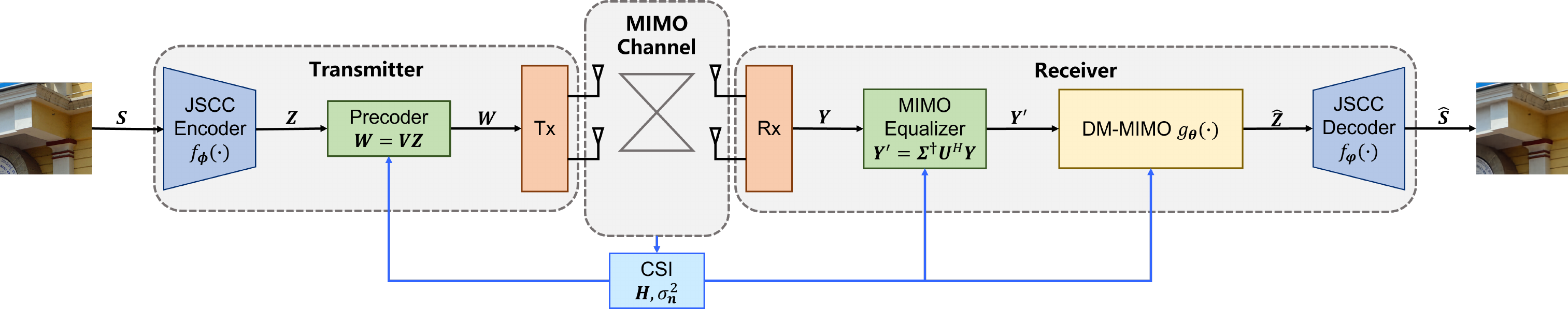}
    \caption{Architecture of DM-MIMO-JSCC.}
    \label{fig:SystemModel}
    \vspace{-1.5em}
\end{figure*}

Inspired by the above, we develop a diffusion model over MIMO channels (DM-MIMO) as a plug-in module at the receiver, eliminating noise and enhancing signal quality through learning signal distribution, thus further improving the performance of semantic communication systems over MIMO channels. Through SVD-based precoding and equalization, MIMO channels are decomposed into parallel sub-channels, each with different effective noise power. 
The existing DMs only consider a fixed noise power in each sampling step, thereby failing to adapt to varying channel conditions across sub-channels.
Given the effective noise power over different sub-channels, DM-MIMO employs different effective sampling steps correspondingly. Based on these effective sampling steps, in order to maintain the correct distribution properties of the input of each sampling step, DM-MIMO applies a joint sampling algorithm, 
adjusting the equalized signal through either noise addition or the reverse sampling process
Moreover, employing a three-stage training algorithm, DM-MIMO is able to learn the distribution of the encoded signal, which enhances the performance of signal denoising and reduces power fluctuations. Utilizing these training and sampling algorithms, the proposed DM-MIMO enhances the robustness of the semantic communication system across a wide range of channel noise power. Additionally, as a plug-in module, DM-MIMO is independent of the structure of JSCC, allowing for flexible implementation in semantic communication systems.

We evaluate the performance of DM-MIMO through extensive experiments. With DM-MIMO, significant reduction of mean square error (MSE) between the decoder input signal and the encoded signal is achieved. This reduction indicates an enhancement in signal quality, thereby enhancing image recovery. As a result, the DM-MIMO semantic communication system (DM-MIMO-JSCC) outperforms the existing JSCC-based semantic communication system in terms of peak signal to noise ratio (PSNR).  



\section{Preliminary of DM and System Overview}\label{System Overview}

\vspace{-0.5em}
\subsection{Preliminary of DM}
DMs achieve data generation through a progressive denoising procedure. With the forward diffusion process gradually corrupting data by adding noise, the reverse sampling process of DMs performs the opposite procedure, producing samples sharing the same distribution as the source data. Specifically, 
for a given data $\bX_0$ with distribution $q(\bX_0)$, the forward diffusion process is derived as a Markov chain, generating a sequence of random variables $\bX_1, \bX_2, \cdots, \bX_T$ modeled by
\begin{equation}
    q\left(\bX_{1:T}|\bX_0\right) \triangleq \prod_{t=1}^T q\left(\bX_t|\bX_{t-1}\right),
\end{equation}
where $T$ is the number of diffusion steps, and $q(\bX_t|\bX_{t-1})$ denotes the conditional distribution of step $t$, formulated as
\begin{equation}
    q(\bX_t|\bX_{t-1}) \triangleq \cN\left(\bX_t; \sqrt{\alpha_t}\bX_{t-1},(1-\alpha_t) \bI\right).
\end{equation}
Here, $\alpha_t \in (0, 1)$ is the noise schedule chosen ahead of model training. 
The reverse sampling process starts by sampling a pure Gaussian noise $\bX_T$ consists of independent and identically distributed (i.i.d.) elements with distribution $\cN(0,1)$, and then gradually generates the target data by 
\begin{equation}
    p_{\btheta} (\bX_0) = p(\bX_T)\prod^T_{t=1} p_{\btheta}(\bX_{t-1}|\bX_t),
\end{equation}
where $\btheta$ are the learnable parameters of DMs, and
\begin{equation}
    p_\btheta(\bX_{t-1}|\bX_t) = \cN\left(\bX_{t-1};\mu_{\btheta}(\bX_t,t),\Sigma_{\btheta}(\bX_t,t)\right).
\end{equation}
Here $\mu_\btheta$ and $\Sigma_\btheta$ denotes the mean and variance of the conditional distribution $p_\btheta(\bX_{t-1}|\bX_t)$ predicted by DMs. As such, DMs are capable of noise elimination. We can first simulate different noise addition process by selecting different diffusion steps, then adopt corresponding sampling steps to recover the source samples. Leveraging the denoising capability of DMs, we propose a plug-in denoising module DM-MIMO, detailed in Section \ref{Methodology}. 

\vspace{-0.5em}
\subsection{System overview}
Consider a semantic communication system performing image transmission over MIMO channels. For simplicity, we assume both the transmitter and receiver are equipped with $M$ antennas. The input image signal is represented by $\bS\in \R^{h\times w\times 3}$, where $h$ and $w$ denote the height and width of the input image respectively, while 3 is the quantity of color channels. The image is transmitted over $k$ channel uses, and the channel bandwidth ratio (CBR) is defined as $\text{CBR} = k/n$, where $n=3hw$.  

As illustrated in Fig. \ref{fig:SystemModel}, at the transmitter, the input image first adopts a JSCC encoding function $f_{\bphi}$ parameterized by $\bphi$ to output the encoded signal $\bZ = [\bz_1, \cdots, \bz_M]^T \in \C^{M\times k}$, expressed as $\bZ = f_{\bphi}(\bS)$. The encoded signal $\bZ$ is then mapped into the channel input signal $\bW \in \C^{M\times k}$ via MIMO precoding. The power constraint $P_s$ is given by $\frac{1}{k}\left\|\bW\right\|^2_F \leq P_s$.

We consider Rayleigh block fading MIMO channels. Let $\bH \in \C^{M\times M}$ be the channel matrix which remains unchanged over $k$ channel uses. Each entry of $\bH$ is i.i.d. random variables following the complex normal distribution $\cCN(0,1)$. Then, the output signal of the MIMO channel can be written as 
\begin{equation}
\label{ChannelModel}
\bY = \bH \bW + \bN,
\end{equation}
where $\bN \in \C^{M\times k}$ is the additive noise term that consists of i.i.d. elements with distribution $\cCN(0, \sigma^2)$, in which $\sigma^2$ denotes the channel noise power. 

As CSI is accessible to both the transmitter and receiver, we adopt SVD-based precoding and equalization to leverage spatial multiplexing and mitigate inter-channel interference. The channel matrix $\bH$ can be decomposed as
\begin{equation}
    \bH=\bU \bSigma \bV ^H, 
\end{equation}
where $\bU = [\bu_1, \cdots, \bu_M] \in \C^{M\times M}$ and $\bV \in \C^{M\times M}$ are 
unitary matrices, and $\bSigma = {\rm diag}[\lambda_1, \cdots, \lambda_M] \in \C^{M\times M}$ 
is a diagonal matrix with singular values $\lambda_1 \geq \cdots \geq \lambda_M$. 
Applying $\bV$ as the precoder at the transmitter, we have $\bW=\bV \bZ$, thus (\ref{ChannelModel}) becomes 
\begin{align}
    \label{receSig}
    \bY = \bH \bV \bZ+ \bN = \bU \bSigma \bZ+\bN.
\end{align}
With the aid of SVD-based equalizer, the equalized signal $\bY '=\bSigma^{\dagger}\bU^H \bY$ at the receiver can be expressed as
\begin{equation}
    \label{equalizedSignal}
    \bY ' = \bZ+\bN ',
\end{equation}
where $\bN ' = \bSigma^{\dagger} \bU^{H} \bN=[\bn '_1, \cdots, \bn '_M]^T$, and
$\bSigma^{\dagger}={\rm diag}[\frac{1}{\lambda_1}, \cdots, \frac{1}{\lambda_M}]$.
Consequently, after employing SVD-based precoding and equalization, the MIMO channels are decomposed into $M$ parallel sub-channels. For sub-channel $i$, the effective noise power is $\sigma_i^2 = \frac{\sigma^2}{\lambda_i^2}$.
In order to remove noise from the equalized signal $\bY '$, we introduce a DM-MIMO module $g_{\btheta}(\cdot)$ with parameters $\btheta$, 
represented as $\bZh=g_{\btheta}(\bY')$, detailed in Section \ref{Methodology}. Finally the JSCC decoding function with parameter $\bvarphi$ takes the denoised signal $\bZh\in \C^{M\times k}$ as input, reconstructing the input image $\bSh=f_{\bvarphi}(\bZh)$.



\vspace{-0.2em}
\section{Design of DM-MIMO} \label{Methodology}

To enhance the robustness of semantic communications, we propose DM-MIMO, a diffusion model eliminating noise over the equalized signal. Due to the high variations of effective noise power over sub-channels, we derive noise-power-aware effective sampling steps and devise a joint sampling algorithm.

\vspace{-0.5em}
\subsection{Analysis of sub-channel conditions}\label{DARS}

Considering block fading, disparities in channel conditions occur over distinct sub-channels. Moreover, under varying channel conditions, the JSCC encoder and decoder jointly learn semantic feature allocation across sub-channels along with semantic feature extraction. The fluctuating channel conditions and the different significance of semantics across parallel sub-channels are crucial factors to be considered in the design of the denoising module.

\begin{figure}
    \centering
    \includegraphics[scale=0.8]{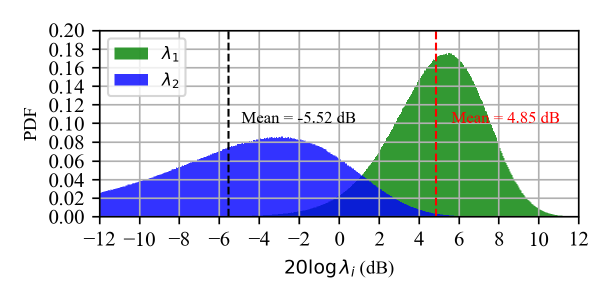}
    \vspace{-1em}
    \caption{Probability density of ${\lambda_i}$ $(i\in \{1,2\})$ in $2\times 2$ MIMO.}
    \label{fig:slow_fading}
    \vspace{-1.5em}
\end{figure}

As $\bU$ is a unitary matrix, the noise of different sub-channels in (\ref{equalizedSignal}) follows an independent Gaussian distribution. Moreover, the effective noise power $\sigma_i^2$ in different sub-channels varies significantly due to its dependence on the channel singular value $\lambda_i$. To illustrate the difference in effective noise power over sub-channels, we perform a Monte Carlo experiment with $1\times 10^7$ samples of $2\times 2$ MIMO Rayleigh fading channels. As shown in Fig. \ref{fig:slow_fading}, the gap between the expectation of $\lambda_i^2$ in the two sub-channels reaches $10.37$ dB, presenting a challenge in joint denoising design. 

 With significant gap in effective noise power between different sub-channels, the naive method of applying one sampling step for all sub-channels fails to eliminate noise effectively. Therefore, an individual effective sampling step need to be employed for each sub-channel according to its effective noise power. Another naive method to remove noise is to employ a separate DM module for each sub-channel. This, however, fails to capture the joint distribution of the encoded signals over different sub-channels. Hence, based on the effective sampling steps, we propose a joint sampling algorithm, which is detailed in the following sub-sections.


\vspace{-0.5em}
\subsection{DM-MIMO} \label{DesignDMMIMO}
We denote the output of each sub-channel as  $\by'_i$, which is the $i$-th row of $\bY'$. 
To adapt to different channel matrices $\bH$, for sub-channel $i$, we employ normalization with factor ${\sigma}_i$ as
\begin{align}
    \bar{\by} '_i &= \frac{1}{\sqrt{1+{{\sigma}_{i}}^2}}\by '_i 
    = \frac{1}{\sqrt{1+{\sigma}_{i}^2}} {\bz_i} + \frac{{\sigma}_{i}}{\sqrt{1+{\sigma}_{i}^2}} \bu^H_i \frac{\bN}{\sigma}.
\end{align}
We define $\bx_{i,0}=\bz_i$, $\bX_0 = [\bx_{1,0}, \cdots, \bx_{M,0}]^T$ and design a forward diffusion process inspired by \cite{song2022denoising} 
\begin{equation}
    \label{forwardDif}
    \bx_{i,t} = \sqrt{\bar{\alpha}_t}\bx_{i,0}+\sqrt{1-\bar{\alpha}_t} \bepsilon_i,
\end{equation}
where $\bar{\alpha}_t = \prod_{l=1}^t \alpha_l$, $t\in \{1, 2, \cdots, T\}$ and $\bepsilon=[\bepsilon_1, \cdots, \bepsilon_M]^T$ denotes the additive noise term consisting of i.i.d. elements following $\cCN(0, 1)$.
To handle signals over sub-channels with different effective noise power, DM-MIMO generates signals sharing the same distribution as $\bar{\by} '_i$. Specifically, DM-MIMO utilizes the forward diffusion process with additive noise power matching the effective noise power, enabling the corresponding sampling process for denoising. Therefore, the effective sampling step $m_i$ is given by
\begin{equation}
    m_i = \mathop{\mathrm{argmin}}\limits_{m\in \{1,2,\cdots,T\}} \left|{\sigma}_{i}^2 - \frac{{1-\bar{\alpha}_{m}}}{{{\bar{\alpha}_{m}}}}\right|.
\end{equation}

By employing tailored effective sampling steps, DM-MIMO is able to handle different signals over different sub-channels in one sampling process, utilizing the inter-sub-channel signal correlations. Moreover, considering the channel noise power varying in a wide range, additive channel noise $\bN'$ introduces dramatic power fluctuations to the equalized signal $\bY'$. DM-MIMO reduces the power fluctuations by learning the distribution of the encoded signal $\mathbf{Z}$ and eliminating noise through the reverse sampling process, thus enhancing the robustness of the receiver.

\begin{figure*}[!t] 
\footnotesize
\vspace{-1em}
\begin{align}
\label{DM-MIMOLoss}
    L =& \, \E \left[-\log {{p_{\btheta}}(\bX_{0}|\bSigma)}\right] \\
    \leq& \, \E_q \left[-\log{\left(\frac{p_{\btheta}(\bX_{0:T},\bY'|\bSigma)}{q(\bX_{1:T},\bY'|\bX_{0}, \bSigma)}\right)}\right] \\
    \label{DM-MIMOLoss-detail}
     =& \, \E_q \Bigg[\underbrace{D_{KL}\left(q\left(\bY '|\bX_{0},\bSigma\right)||p\left(\bY'|\bSigma\right)\right)}_{L_{\bY '}}-\underbrace{\log{p_{\btheta} \left(\bX_{0}|\bX_{1},\bSigma\right)}}_{L_{0}} \nonumber 
     + \underbrace{D_{KL}\left(q\left(\bX_{T}|\bY', \bX_0, \bSigma\right) || p\left(\bX_{T}|\bY', \bSigma\right)\right)}_{L_{T}} \nonumber \\
     & + \sum^{T}_{t=1} \underbrace{{D_{KL}\left(q\left(\bX_{t-1}|\bX_{t}, \bX_{0}, \bSigma\right) || p_{\btheta}\left(\bX_{t-1}|\bX_{t}, \bSigma\right)\right)}}_{L_{t-1}}\Bigg].
\end{align} 
\hrule
\vspace{-1em}
\end{figure*}

In the training process, as $\bx_{i,m_i}$ and $\by^{\prime}_{i}$ share the same distribution,
DM-MIMO can be trained on the forward diffusion process of $\bZ$ instead of $\bY^{\prime}$. Aiming to recover ${\bZ}$ through learning its distribution, the loss function $L$ is defined by the variational bound on the negative log likelihood function of $\bX_{0}$,
as given in (\ref{DM-MIMOLoss}).
Analyzing the components of loss function, $L_{\bY'}$ and $L_{t}$ can be ignored during training as they are not related to ${\btheta}$. Therefore, we focus on $L_{0}$ and $L_{t-1}$, revealing the training goal of approximating the distribution of $q(\bX_{t-1}|\bX_{t},\bX_{0},\bSigma)$ with $p_{\btheta}(\bX_{t-1}|\bX_{t}, \bSigma)$.
After re-parameterization and re-weighting, the loss function $L_{t-1}$ can be simplified as
\begin{equation}
    \label{L_t-1}
    \E_{\bX_{0}, \bepsilon, \bSigma} \left ( \left\|\bepsilon - \bepsilon_{\btheta}(\bX_t, \bSigma, t)\right\|_2^2 \right).
\end{equation}
Finally, to optimize (\ref{L_t-1}) for all $t\in \{1,\cdots, T\}$, the loss function of DM-MIMO can be expressed as
\begin{equation}
    L_{DM-MIMO}({\btheta})= \E_{\bX_{0}, \bepsilon, \bSigma, t}\left(\left\|\bepsilon - \bepsilon_{\btheta}(\bX_t, \bSigma, t)\right\|_2^2\right).
\end{equation}


The training process of DM-MIMO is detailed in Algorithm \ref{TADM}. With different effective sampling steps chosen according to different effective noise power $\sigma_i^2$, DM-MIMO demonstrates the ability of denoising under diverse channel conditions while utilizing fixed parameters.

\begin{algorithm}[t]
    \caption{Training Algorithm of DM-MIMO}
    \hspace*{0.02in} 
    \small{\bf{Input:}}
    \small{Encoded signal set, diffusion steps $T$ and noise schedule parameter $\bar{\alpha}_t$ for $t\in\{1, \cdots, T\}$; } \\
    \hspace*{0.02in} 
    \small{\bf{Output:}}
    \small{Trained DM-MIMO model parameters $\btheta$; }
    \label{TADM}
    \begin{algorithmic}[1]
        \While {the stop condition is not met}
          \State Randomly sample $\bZ$ from encoded signal set
          \State Randomly sample $t$ from $Uniform(\{1,\cdots,T\})$
          \State Randomly sample $\bH$ 
          \ForAll {$i=1$ to $M$} 
            \State Randomly sample $\bepsilon_i$ from $\cN\left(0, \bI_{2k}\right)$ 
          \EndFor
        \State $\bepsilon = [\bepsilon_1, \cdots, \bepsilon_M]$
        \State Generate sample $\bX_{t} = \sqrt{\bar{\alpha}_t} \bZ + \sqrt{1-\bar{\alpha}_t} \bepsilon$ 
        \State Take gradient descent step: $\nabla_{\btheta} \left(\left\|\bepsilon -\bepsilon_{\btheta} \left(\bX_{t}, \bSigma, t\right)\right\|_2^2\right)$
        \EndWhile
 \end{algorithmic}
\end{algorithm}

\vspace{-0.5em}
\subsection{Sampling Algorithm of DM-MIMO}
 Inspired by \cite{Lugmayr_2022_CVPR}, we design a joint sampling algorithm for DM-MIMO. 
 Specifically, in the $t$-th sampling step of DM-MIMO, in order to guide the denoising process of equalized signals with high effective noise power, we add noise to the equalized signals with low effective noise power and send them to the reverse sampling step. In a word, 
 we employ either noise addition or the reverse sampling process for each sub-channel based on the value of its effective sampling step $m_i$.

\begin{figure}[t]
    \vspace{-1em}
    \centering
    \includegraphics[scale=0.25]{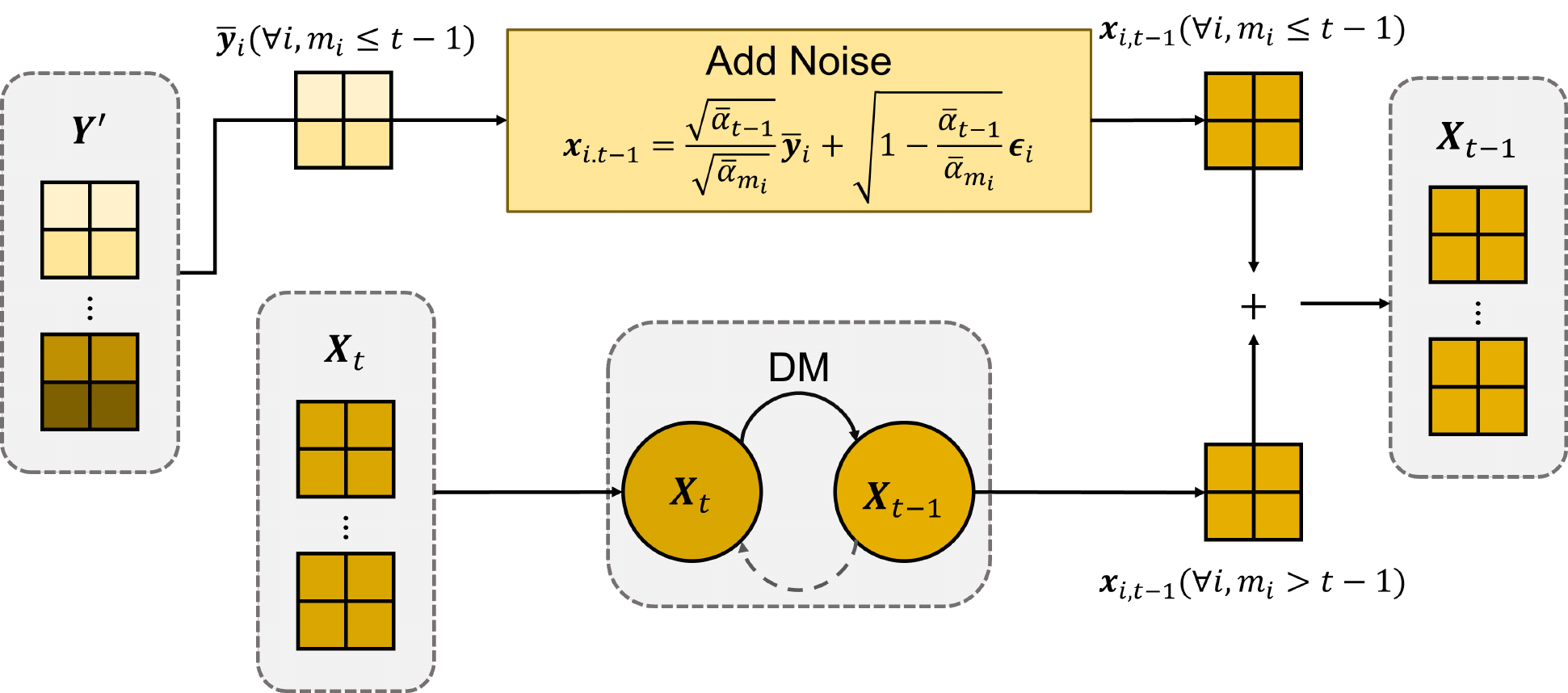}
    \caption{The $t$-th sample step of DM-MIMO joint sampling algorithm.}
    \label{fig:DMMIMO_detail}
    \vspace{-1.5em}
\end{figure}
We begin the sampling process from sampling step $t=\max\{m_1, \cdots, m_M\}$. As illustrated in Fig. \ref{fig:DMMIMO_detail}, 
if $m_i \leq (t-1)$, to keep the correct properties of the distribution of $\bX_{t-1}$, we derive $\bx_{i,t-1}$ by adding noise to $\bar{\by}'_i$, given by
\begin{align}
    \bx_{i,t-1} = \sqrt{\frac{\bar{\alpha}_{t-1}}{\bar{\alpha}_{m_i}}}\bar{\by}'_i + \sqrt{1-\frac{\bar{\alpha}_{t-1}}{\bar{\alpha}_{m_i}}}\bepsilon_i.
\end{align}
On the other hand, if $m_i > (t-1)$, assume the knowledge of $\bx_{i,t}$ and $\bx_{i,0}$ to be available, we can derive the sampling process of $\bx_{i,t-1}$ as
\begin{equation}
    \bx_{i,t-1}=\sqrt{\bar{\alpha}_{t-1}}\bx_{i,0}+\sqrt{1-\bar{\alpha}_{t-1}}\frac{\bx_{i,t}-\sqrt{\bar{\alpha}_t}\bx_{i,0}}{\sqrt{1-\bar{\alpha}_t}},
\end{equation}
where $\bx_{i,0}$ can be acquired by re-writing (\ref{forwardDif}) as
\begin{equation}
    \bx_{i,0} = \frac{1}{\sqrt{\bar{\alpha}_t}}\left(\bx_{i,t}-\sqrt{1-\bar{\alpha}_t} \bepsilon_i\right).
\end{equation}
In the reverse sampling process, with only $\bx_{i,t}$ and $\bepsilon_{\btheta, i}(\bX_{t}, \bSigma, t)$ known to the receiver, $\bepsilon_i$ is replaced with $\bepsilon_{\btheta, i}(\bX_{t}, \bSigma, t)$. Therefore, the sampling process can be expressed as
\begin{align}
    \bx_{i,t-1}=&\, \sqrt{\bar{\alpha}_{t-1}}\left(\frac{1}{\sqrt{\bar{\alpha}_t}}\left(\bx_{i,t}-\sqrt{1-\bar{\alpha}_t} \bepsilon_{\btheta, i}\left(\bX_{t}, \bSigma, t\right)\right)\right) \nonumber \\
    &+\sqrt{1-\bar{\alpha}_{t-1}} \bepsilon_{\btheta, i}\left(\bX_{t}, \bSigma, t\right).
\end{align}

As for the last sampling step $t=1$, DM-MIMO only predicts $\bx_{i,0}$ with $\bx_{i,1}$, given by
\begin{equation}
    \bx_{i,0} = \frac{1}{\sqrt{\bar{\alpha}_1}}\left(\bx_{i,1}-\sqrt{1-\bar{\alpha}_1}  \bepsilon_{\btheta, i}\left(\bX_{1}, \bSigma, 1\right)\right).
\end{equation}

As outlined in Algorithm \ref{SADM}, for sub-channel $i$, by comparing $m_i$ with the sampling step $t$, the joint sampling algorithm either adds noise or performs the reverse sampling process, thus leveraging all the semantic information while addressing different effective noise power over different sub-channels. 

\begin{algorithm}[t]
    \caption{Sampling Algorithm of DM-MIMO}
    \hspace*{0.02in} 
    \small{\bf{Input:}}
    \small{Equalized signal $\bY '$, channel matrix $\bH$, channel noise power $\sigma^2$;} \\
    \hspace*{0.02in} 
    \small{\bf{Output:}}
    \small{Denoised signal $\hat{\bZ}$; }
    \label{SADM}
    \begin{algorithmic}[1]
        \State $[\by '_1, \cdots, \by '_M] = \bY '$
        
        \ForAll {$i=1$ to $M$}
            \State Calculate $\lambda_i$ and $\sigma_i$ based on $\bH$ and $\sigma^2$
            \State $\bar{\by} '_i = \frac{1}{\sqrt{1+{{\sigma}_i}^2}}\by '_i$
            \State Calculate $m_i$ according to sub-channel ${\sigma}_i$
        \EndFor
        \State $m_{max} = \max\{m_1, \cdots, m_M\}$
        \State $t = m_{max}$
        \ForAll{$i=1$ to $M$}
            \State Randomly sample $\bepsilon_i$ from $\cN\left(0, \bI_{2k}\right)$ 
            \State $\bx_{i,t} = \sqrt{\frac{\bar{\alpha}_{t}}{\bar{\alpha}_{m_i}}}\bar{\by}'_i+\sqrt{1-\frac{\bar{\alpha}_{t}}{\bar{\alpha}_{m_i}}}\bepsilon_i$
        \EndFor
        \For {$t=m_{max}, m_{max}-1,\cdots,2$}
            \State $\bX_t = [\bx_{1,t}, \cdots, \bx_{M,t}]$
            \ForAll {$i=1$ to $M$}
                \If {$ m_i\leq t-1$}
                    \State Randomly sample $\bepsilon_i$ from $\cN\left(0, \bI_{2k}\right)$ 
                    \State $\bx_{i,t-1} = \sqrt{\frac{\bar{\alpha}_{t-1}}{\bar{\alpha}_{m_i}}}\bar{\by}'_i+\sqrt{1-\frac{\bar{\alpha}_{t-1}}{\bar{\alpha}_{m_i}}}\bepsilon_i$
                \Else
                    \State $\hat{\bepsilon}_{i}= \bepsilon_{\btheta, i}\left(\bX_t, \bSigma, t\right)$
                    \State $\bx_{i,t-1} = \sqrt{\bar{\alpha}_{t-1}}\left(\frac{\bx_{i,t} - \sqrt{1-\bar{\alpha}_t}\hat{\bepsilon}_{i}}{\sqrt{\bar{\alpha}_t}}\right)+\sqrt{1-\bar{\alpha}_{t-1}}\hat{\bepsilon}_{i}$
                \EndIf
            \EndFor
        \EndFor
        \State $t=1$
        \State $\hat{\bepsilon}= \bepsilon_{\btheta}\left(\bX_1, \bSigma, t\right)$
        \State $\bZh = \frac{\bX_1 - \sqrt{1-\bar{\alpha}_1}\hat{\bepsilon}}{\sqrt{\bar{\alpha}_1}}$
 \end{algorithmic}
\end{algorithm}

\vspace{-0.5em}
\subsection{Training algorithm of semantic communication system}

Since DM-MIMO is trained to learn the distribution of the encoded signal $\bZ$, a three-stage training algorithm is proposed. In the first stage, the JSCC encoder and decoder 
are jointly trained to minimize the reconstruction distortion.
With MSE adopted as performance metric, the loss function of the first training stage can be written as 
\begin{equation}
    L_{s1}\left(\bphi, \bvarphi\right) = \E_{\bS \sim p_\bS}\E_{\bY \sim p_{\bY|\bS}}\left \|\bS-\hat{\bS}\right\|^2_F.
\end{equation}

With the well-trained and fixed parameters of the semantic encoder, DM-MIMO is trained in the second stage using Algorithm \ref{TADM}. Specifically, DM-MIMO adopts the whole encoded signal $\bZ$ as input, leveraging all the semantic information received. Benefiting from the different effective sampling steps and the joint sampling algorithm, DM-MIMO is capable of noise elimination over different sub-channels. 

In the third stage, the JSCC decoder is retrained to adapt to the JSCC encoder and the DM-MIMO. Though only the parameters of the JSCC decoder are trained, the entire system operates under real MIMO channels.
The loss function is
\begin{equation}
    L_{s3}\left(\bvarphi\right) = \E_{\bY' \sim p_{\bY'|\bS}}\left\|\bS-\bSh\right\|^2_F.
\end{equation}

To investigate the robustness of semantic communication systems, we consider a universal transmission scenario with random transmit power. Therefore, to enhance robustness of the communication system, the channel noise power is randomly chosen from a regime during both the first and the third training stage.

\vspace{-0.2em}
\section{Experimental results} \label{Experimental results}

In this section, a series of experiments are conducted to evaluate the performance of DM-MIMO-JSCC. 

\vspace{-0.5em}
\subsection{Experiment Setup}
We consider DIV2K dataset in the experiments. This dataset contains $1000$ diverse 2K images from a wide range of real-world scenes, $800$ of which are used for training, $100$ for validating and the rest $100$ for testing. We randomly crop the images into $256\times 256$ patches in the training process. We consider block-fading MIMO channels with antenna number $M=2$ and channel SNRs ranging from $0$ dB to $20$ dB, where the channel SNR is defined as 
\begin{align}
    \label{ChanSNR}
    SNR = 10\log_{10}\frac{\E_{\bH, \bW}\left[\left\|\bH \bW\right\|_F^2\right]}
    {\E_\bN\left\|\bN\right\|_F^2} = 10\log_{10}\frac{P_s}{\sigma^2}.
\end{align}

We adopt the latest Swin-Transformer based JSCC \cite{10094735} for the JSCC in our DM-MIMO-JSCC. For simplicity, channel adaptation is not considered in the JSCC encoder and decoder.
The proposed DM-MIMO is established on U-Net architecture \cite{10.1007/978-3-319-24574-4_28}. We choose hyper-parameter $T=1000$ and employ a noise schedule $\alpha_t$ that linearly decreases from $\alpha_1=0.9999$ to $\alpha_T=0.98$. 
DM-MIMO is trained with an Adam optimizer for 800 epochs, which adopts a cosine warm-up learning rate schedule with initial learning rate of $1\times 10^{-4}$. Besides, the JSCC is trained for $800$ epochs during the first training stage, with a learning rate of $1\times 10^{-4}$. The retraining epochs in the third training stage is set to $20$. We implement DM-MIMO-JSCC on one NVIDIA A40 GPU using Pytorch.
For fair comparison, the same Swin-Transformer-based JSCC but without DM-MIMO is considered as a benchmark. It is trained with the same setup as the first training stage of DM-MIMO-JSCC. 

\begin{figure}[t]
    \vspace{-1em}
    \centering
		\includegraphics[width=0.85\linewidth, trim = 5 5 5 5,clip]{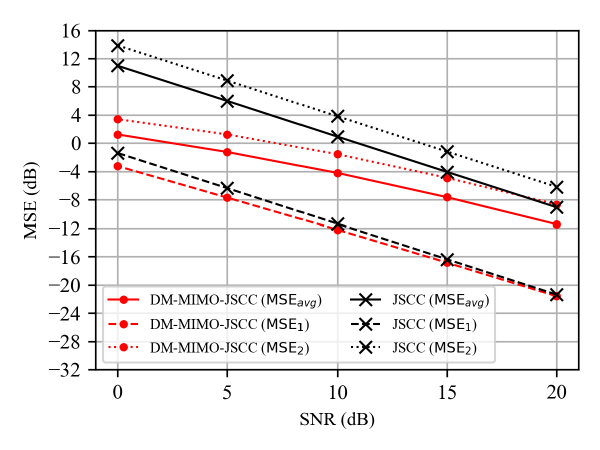}
        \vspace{-0.3em}
		\caption{MSE performance versus SNR under $2\times2$ MIMO channel. CBR is set to $1/128$.}
        \label{fig:DIV24_MSE}
    \vspace{-1.5em}
\end{figure}

\begin{figure}[t]
    \centering
		\includegraphics[width=0.85\linewidth, trim = 5 5 5 5,clip]{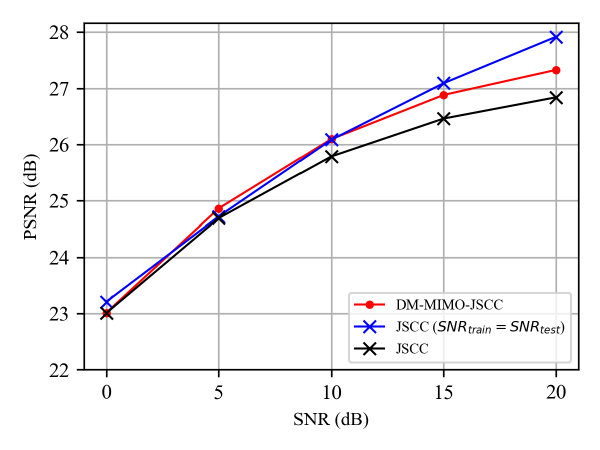}
        \vspace{-0.3em}
		\caption{PSNR performance versus SNR under $2\times2$ MIMO channel. CBR is set to $1/128$.}
        \label{fig:DIV24_PSNR}
    \vspace{-1.5em}
\end{figure}

\vspace{-0.5em}
\subsection{MSE performance}

We first evaluate the effectiveness of DM-MIMO by comparing the MSE between the denoised signal and the encoded signal in DM-MIMO-JSCC and the MSE between the equalized signal and the encoded signal in universal JSCC without DM-MIMO. ${\rm MSE}_{avg}$ denotes the average MSE across all sub-channels, while ${\rm MSE}_{i}$ denotes the average MSE of the $i$-th sub-channel. As shown in Fig. \ref{fig:DIV24_MSE}, with DM-MIMO adopted, both ${\rm MSE}_1$ and ${\rm MSE}_2$ decrease over the SNR regime of $[0, 20]$ dB. The lower the channel SNR, the higher the MSE gain achieved by DM-MIMO on both sub-channels. The MSE gain of ${\rm MSE}_1$ and ${\rm MSE}_2$ achieve over $0.233$ dB and $2.482$ dB respectively in the SNR regime of $[0, 20]$ dB. As such, our proposed DM-MIMO, with effective sampling step adaptation and joint sampling algorithm, demonstrates effectiveness in noise elimination and signal quality enhancement.

\vspace{-0.5em}
\subsection{PSNR performance}
\begin{figure}[t]
    \centering
    \includegraphics[width=0.95\linewidth, trim = 15 15 25 10,clip]{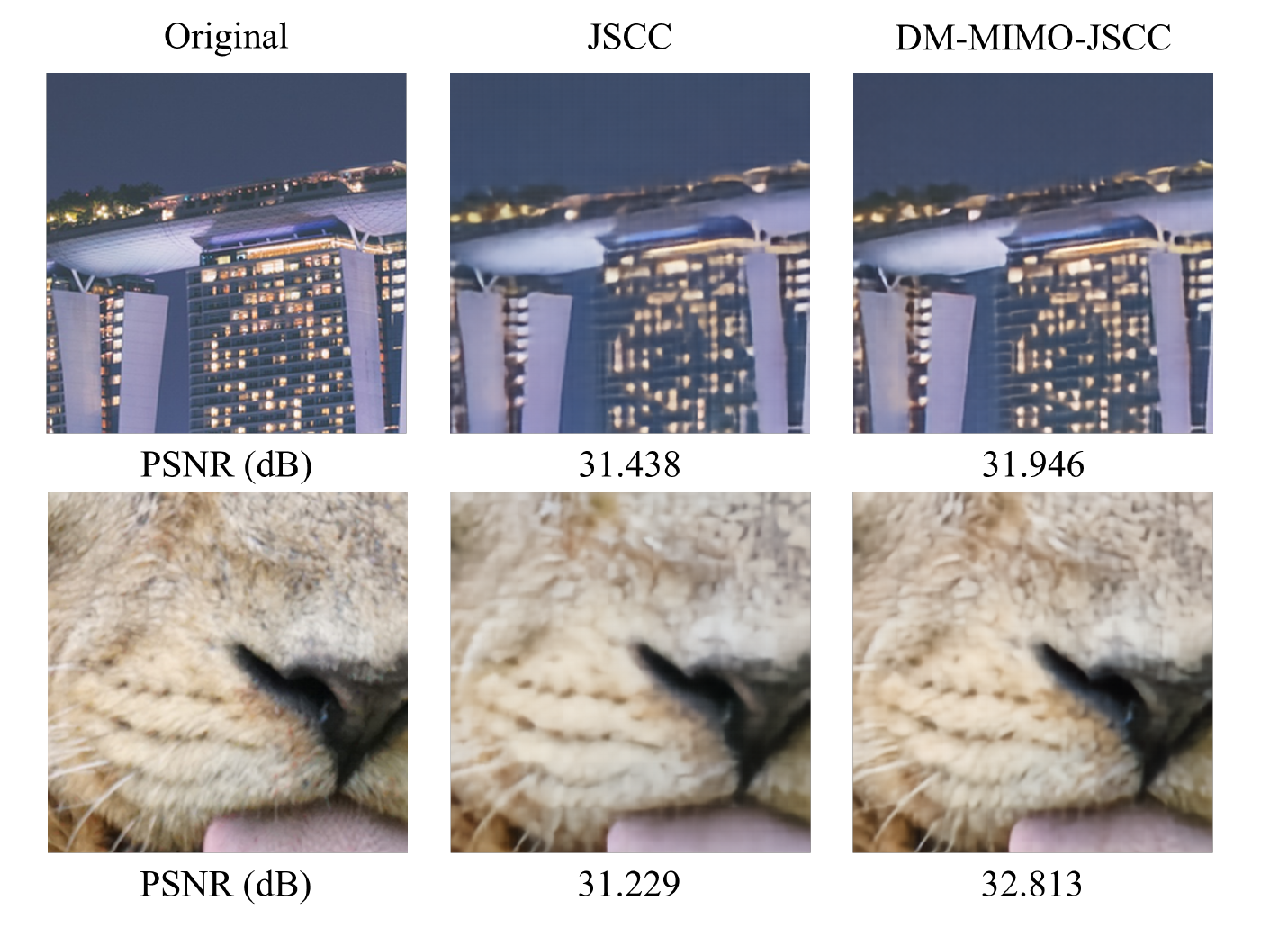}
    \vspace{-0.3em}
    \caption{Examples of visualization results under $2\times2$ MIMO channel with channel SNR of $20$ dB.}
    \label{fig:Sample}
    \vspace{-1.5em}
\end{figure}

Fig. \ref{fig:DIV24_PSNR} shows the PSNR performance versus the channel SNR. Our DM-MIMO-JSCC outperforms universal JSCC without DM-MIMO in SNR regime from $0$ dB to $20$ dB. The higher the channel SNR, the higher the PSNR gain achieved by DM-MIMO-JSCC. Specifically, DM-MIMO-JSCC achieves a PSNR gain of $0.488$ dB at SNR = $20$ dB. 
For comparison, the JSCC scheme where training SNR matches testing SNR is also ploted in Fig. \ref{fig:DIV24_PSNR}. It can be seen that the DM-MIMO-JSCC achieves comparable performance in SNR regime of $[0, 15]$ dB. Moreover, we compare the reconstructed samples of different methods under channel SNR of $20$ dB. As shown in Fig. \ref{fig:Sample}, the samples reconstructed by DM-MIMO-JSCC show better visual quality compared with those reconstructed by universal JSCC without DM-MIMO, as the first one shows a sharper edge and the second one shows a clearer fur detail. In a word, by eliminating noise of the equalized signal, DM-MIMO enhances the robustness of the semantic communication system and achieves better performance in image recovery over various channel conditions.

\vspace{-0.5em}
\subsection{Complexity Analysis}
We analyze the computational complexity of the proposed DM-MIMO in terms of multiply accumulate operations (MACs). Table \ref{tab:Complex} shows the number of MACs for one step of sampling in DM-MIMO at different CBRs. It can be seen that the amount of MACs of DM-MIMO increases proportionally with the square of CBR. This is expected as the number of channels in U-Net adopted by DM-MIMO directly depends on the length of the received signal. Meanwhile, the number of MACs in JSCC only increases slightly when CBR increases. As such, our DM-MIMO is more suitable at lower CBR region in terms of computational complexity.

\section{Conclusion} \label{Conclusion}
In this paper, we propose a plug-in channel denoising module named DM-MIMO, aiming at enhancing the robustness of semantic communication systems over MIMO channels. By learning the distribution of the encoded signal, DM-MIMO eliminates noise and reduces power fluctuations of the decoder input signal, thereby enhancing the robustness of the semantic communication system. To address the diversity of sub-channel conditions, we employ effective sampling steps correspondingly, and devise a joint sampling algorithm to leverage all the received semantic information while managing the variations in effective noise power. 
Experimental results demonstrate that DM-MIMO-JSCC outperforms JSCC without DM-MIMO in image recovery.
\begin{table}[t]
\caption{MACs of DM-MIMO and JSCC with different CBRs.}
\setlength{\tabcolsep}{0.08\linewidth}
\begin{center}
\vspace{-1em}
\begin{tabular}{c|c|c}
\Xhline{1pt}
{CBR} & {DM-MIMO}& {JSCC} \\
\hline
\hline
0.0026 & 6.762 G & 32.723 G \\
\hline
0.0039 & 15.215 G & 32.726 G \\
\hline
0.0078 & 60.858 G & 32.736 G \\
\hline
0.0104 & 108.193 G & 32.742 G \\
\Xhline{1pt}
\end{tabular}
\label{tab:Complex}
\end{center}
\vspace{-2.5em}
\end{table}

\bibliographystyle{IEEEtran}
\bibliography{main}


\end{document}